\documentclass[notitlepage,superscriptaddress,12pt]{revtex4}
\usepackage{graphicx}
\usepackage{hyperref}

\begin{document}

\title{Local quantum Fisher information and
one-way quantum deficit in spin-$\frac{1}{2}$
$XX$ Heisenberg chain with three-spin interaction}
\author{Biao-Liang Ye}
\email{biaoliangye@gmail.com}
\affiliation{Quantum Information Research
Center, Shangrao Normal
University, Shangrao 334001, China}
\author{Bo Li}
\affiliation{School of Mathematics $\&$
Computer Science, Shangrao Normal
University, Shangrao 334001, China}
\affiliation{Quantum Information Research
Center, Shangrao Normal
University, Shangrao 334001, China}
\author{Xiao-Bin Liang}
\affiliation{School of Mathematics $\&$
Computer Science, Shangrao Normal
University, Shangrao 334001, China}
\affiliation{Quantum Information Research
Center, Shangrao Normal
University, Shangrao 334001, China}
\author{Shao-Ming Fei}
\email{feishm@cnu.edu.cn}
\affiliation{School of Mathematical Sciences,
Capital Normal University, Beijing 100048,
China}
\affiliation{Max-Planck-Institute for
Mathematics in the Sciences, 04103 Leipzig, Germany}
\date{\today}

\begin{abstract}
We explore quantum phase transitions in the
spin-1/2 $XX$ chain with three-spin interaction in terms of 
	local quantum Fisher information
	and one-way quantum deficit, together with the
demonstration of quantum fluctuations.
Analytical results are derived and analyzed in detail.
\end{abstract}
\maketitle

\section{Introduction}
Quantum entanglement plays a vital role in quantum information
processing \cite{Horodecki2009}.
As an important resource, quantum entangled states have been used in
quantum teleportation \cite{Bennett1993},
remote state preparation \cite{Bennett2001},
secure quantum-communications network \cite{Yin2017},
etc. Besides quantum entanglement, quantum discord
characterizes non-classical correlations \cite{Modi2012}.
The one-way quantum deficit \cite{Ye2016} is another
key measure to describe quantum correlation \cite{Streltsov2011}.
While the quantum Fisher information \cite{Petz2002,Ye2018b} is important in
the estimation accuracy scenarios.

On the other hand, the quantum phase transitions have received much
attention in condensed matter physics \cite{Sachdev1999}.
The quantum fluctuations are able to be illustrated by quantum correlations.
In Ref. \cite{Osterloh2002} the role of entanglement played in phase transition and theory of critical
phenomena in $XY$ system has been investigated.
The quantum discord and entanglement between the nearest-neighbor qubits
in an infinite (spin-1/2) chain described by the Heisenberg model
(XXZ Hamiltonian) were investigated in \cite{Werlang2010}, and the critical points
associated with quantum phase transitions at finite temperature had been analyzed.
In Ref.\cite{Liu2011}, the authors studied the effects of
Dzyaloshinskii-Moriya interaction on pairwise quantum discord, entanglement, and
classical correlation in the anisotropic $XY$ spin-half chain.
It has been shown that the quantum discord can be useful to highlight the
quantum phase transition, especially for the long-distance spins,
while entanglement decays rapidly.
The quantum discord has been also used 
to show the quantum phase transition
in $XX$ model in \cite{BenQiong2011}.
In Ref. \cite{Li2016} the authors connected the local quantum coherence
based on Wigner-Yanase skew information and the quantum phase transitions.
The local quantum coherence and its derivatives are used effectively in
detecting different types of quantum phase transitions in different spin systems.
In Ref. \cite{Chen2016} the authors introduced a coherence susceptibility
approach in identifying quantum phase transitions induced by quantum fluctuations.

Although different measures of quantum correlations have
been used to characterize quantum phase
transitions in different spin chain systems,
both local quantum Fisher information
and one-way quantum deficit have not been adopted to
study quantum phase transition in Hersenberg
$XX$ models. In this paper, we investigate the quantum
phase transitions of the $XX$ chain with
three spin interaction. We explore the quantum
fluctuation via both local quantum Fisher information
and one-way quantum deficit to investigate the quantum phase transition.
We review the basic definitions of local quantum Fisher
information and one-way quantum deficit in Sec. II.
In Sec.III, the Hersenberg spin-$\frac12$
is introduced and the main results are presented.
Conclusions are given in Sec. IV.

\section{Preliminaries}
We first recall the basic definitions of local quantum Fisher
information and one-way quantum deficit.

{\sf Local quantum Fisher information}~
Quantum Fisher information (QFI) is recognized
as the most widely used quantity for
characterizing the ultimate accuracy in
parameter estimation scenarios.
Recently, many efforts have been made
toward evaluating the dynamics of QFI to
establish the relevance of quantum
entanglement in quantum metrology.
It has been demonstrated that quantum entanglement
leads to a notable improvement of the accuracy of parameter estimation.
It is natural to ask whether quantum correlations
beyond quantum entanglement can
be related to the precision in quantum metrology protocols.

Generally, for an arbitrary quantum state $\rho_{\theta}$
that depends on a 
variate $\theta$,
the QFI is expressed by \cite{Slaoui2019},
\begin{eqnarray}
	F(\rho_{\theta})=\frac{1}{4}
	{\rm Tr}[\rho_{\theta}L_{\theta}^2].
\end{eqnarray}
Here the symmetric logarithmic derivative
$L_{{\theta}}$ is get as the solution
of the following equation
\begin{eqnarray}
	\frac{\partial{\rho_{\theta}}}{\partial\theta}=\frac12(L_{\theta}\rho_\theta+
	\rho_\theta L_\theta).
\end{eqnarray}
The parametric states $\rho_{{\theta}}$
can be derived from an initial probe
state $\rho$ subjected to a unitary
transformation $U_{\theta}=e^{-i H\theta}$ which is
dependent on $\theta$ and a Hermitian operator $H$, i.e., $\rho_{\theta}=U_{\theta}^{\dagger}\rho U_{{\theta}}$.
Thus $F(\rho_{\theta})$ is given by
\begin{equation}
	F(\rho,H)=\frac12\sum_{i\ne j}
	\frac{(p_i-p_j)^2}{p_i+p_j}|\langle
	\psi_i|H|\psi_j\rangle|^2,
\end{equation}
where $\rho=\sum_{{i=1}}p_{i}|\psi_i\rangle\langle\psi_i|$ is spectral decomposition of $\rho$,
$p_{i}\ge0$ and $\sum_{{i=1}}p_{i}=1$.

Now consider a bipartite quantum state $\rho_{{AB}}$ in
the Hilbert space $W=W_{A}\otimes W_{{B}}$.
We assume that the dynamics of the first
subsystem is subjected by the local phase shift
transformation $e^{{-i\theta H_A}}$, with
$H_{A}=H_{a}\otimes I_{B}$ the local
Hamiltonian. Therefore QFI reduces to
local quantum Fisher information (LQFI),
\begin{eqnarray}\label{qfi}
F(\rho, H_A)={\rm Tr}(\rho H_{A}^2)-
\sum_{i\ne j}\frac{2p_ip_j}{p_i+p_j}
|\langle\psi_i|H_A|\psi_j\rangle|^2. 	
\end{eqnarray}

Local quantum Fisher information was introduced
to deal with pairwise quantum measures
of discord type. The local
quantum Fisher information $Q(\rho)$ is
defined as the minimum quantum Fisher
information over all local Hamiltonians
$H_{A}$ acting on the subsystem $A$ \cite{Kim2018},
\begin{equation}
	Q(\rho)=\min_{H_A}F(\rho,H_A).
\end{equation}
For local Hamiltonian $H_{a}=\vec{\sigma}\cdot\vec{r}$,
with $|\vec{r}|=1$ and $\vec{\sigma}=(\sigma_x,\sigma_y,\sigma_z)$
the Pauli matrices, one has
${\rm Tr}(\rho H_A^2)=1$, and the second
term in (\ref{qfi}) can be written as
\begin{eqnarray}
	\sum_{i\ne j}\frac{2p_ip_j}{p_i+p_j}
	|\psi_i|H_A|\psi_j\rangle|^2=
	\sum_{i\ne j}\sum_{l,k=1}^{3}
	\frac{2p_ip_j}{p_i+p_j}\langle\psi_i
	|\sigma_l\otimes I_B|\psi_j\rangle
	\langle\psi_j|\sigma_k\otimes I_B|\psi_i
	\rangle
	=\vec{r}^\dagger\cdot T\cdot \vec{r},
\end{eqnarray}
where the elements of the $3\times 3$
symmetric matrix $T$ are given by
\begin{equation}\label{el}
	T_{lk}=\sum_{i\ne j}\frac{2p_ip_j}{p_i+p_j}\langle\psi_i|\sigma_l\otimes I_B|\psi_j\rangle\langle\psi_j|\sigma_k\otimes I_B|
	\psi_i\rangle.
\end{equation}

To minimize $F(\rho,H_A)$, it is necessary
to maximize the quantity $\vec{r}^{\dagger}\cdot T\cdot \vec{r}$ over all unit vectors
$\vec{r}$. The maximum value coincides with
the maximum eigenvalue of $T$. Hence, the
minimal value of local quantum Fisher
information $Q(\rho)$ is
\begin{equation}\label{lqfi}
	Q(\rho)=1-\lambda_{\max}(T),
\end{equation}
where $\lambda_{\max}$ denotes the
maximal eigenvalue of the symmetric matrix
$T$ defined by (\ref{el}).

{\sf One-way quantum deficit (OWQD)}~
The one-way quantum deficit is defined
as the difference of the von Neumann entropy of a bipartite state, $\rho_{AB}$,
before and after a measurement performed on,
without a loss of generality, subsystem $B$ \cite{Streltsov2011},
\begin{equation}\label{deficit}
	\Delta=\min_{\Pi_A^i} S(\sum_i\Pi_k^B(\rho_{AB}))-S(\rho_{AB}),
\end{equation}
where $\Pi_k^B$ is the measurement on
subsystem $B$ and $S(\rho_{AB})=-{\rm Tr}\rho_{AB} \log\rho_{AB}$ is the
von Neumann entropy. Throughout the article, $\log$ is in base 2. The minimum is taken over
all local measurements $\Pi_k^B.$

The states after the projective measurement can be expressed as
\begin{eqnarray}
	\widetilde{\rho}_{AB}=
	\sum_{k}^{}(I\otimes\Pi_k)\rho_{AB}
	(I\otimes\Pi_k)^\dagger.
\end{eqnarray}
The post-measurement states is given by
\begin{eqnarray}
	\rho_{AB}^k=\frac{1}{p_k}
	(I\otimes\Pi_k)\rho_{AB}
	(I\otimes\Pi_k)^\dagger,
\end{eqnarray}
where
\begin{eqnarray}
	p_k={\rm Tr}[
	(I\otimes\Pi_k)\rho_{AB}
	(I\otimes\Pi_k)^\dagger].
\end{eqnarray}
Here $\Pi_{k}~(k=0,1)$ are the
general orthogonal projectors
\begin{eqnarray}
	\Pi_k=V|k\rangle\langle k|V^\dagger,
\end{eqnarray}
where ${V}$ belongs to the special
unitary group $SU(2)$. The rotations $V$
may be parametrized by two parameters
$\theta$ and $\phi$, respectively,
\begin{eqnarray}
	V=\left(
	\begin{array}{cc}
		\cos(\theta/2) & -e^{-i\phi}\sin(\theta/2)\cr
		e^{i\phi}\sin(\theta/2) & \cos(\theta/2)
	\end{array}\right),
\end{eqnarray}
with $\theta\in[0,\pi]$ and $\phi\in[0,2\pi]$.

\section{Quantum phase transition
of the Heisenberg-$\frac12$ $XX$ spin chain
model with three spin interaction}
The Hamiltonian of the Heisenberg spin-$\frac12$ $XX$ chain
can be given as \cite{Lou2004},
\begin{eqnarray}\label{ha}
	H&=&\sum_{l=1}^{N}-J(S_l^xS_{l+1}^x
	+S_l^yS_{l+1}^y+\Delta S_l^zS_{l+1}^z)
	-J'\{(S_{l-1}^xS_l^zS_{l+1}^y
	-S_{l-1}^yS_l^zS_{l+1}^x)\nonumber\\
	&+&\Delta(S_{l-1}^yS_l^xS_{l+1}^z
	-S_{l-1}^zS_l^xS_{l+1}^y)
	+\Delta(S_{l-1}^zS_l^yS_{l+1}^x
	-S_{l-1}^xS_l^yS_{l+1}^z)\}.
\end{eqnarray}
Here  $N$ is the total number of spins, $S_{l}^q~ (q=x,y,z)$ are
spin operators of $S=1/2$-spin on site
$l$,
$J$ is the nearest-neighbor Heisenberg
exchange coupling, $J'$ is the strength of three
spin interaction, and $\Delta$
denotes the anisotropy parameter. The model
shows several quantum phases depending
on the parameters $\Delta$ and $J'/J$.
The same Hamiltonian is used to investigate the
 current-carrying states for the system
with only the nearest-neighbor interactions,
where the three-spin terms play the role of
the Lagrange multiplier.

If $\Delta=0$, the Hamiltonian (\ref{ha})
reduces to a free spinless fermion model,
\begin{eqnarray}
	H=\sum_{l=1}^{N}-J(S_l^xS_{l+1}^x
	+S_l^yS_{l+1}^y)
	-J'\{(S_{l-1}^xS_l^zS_{l+1}^y
	-S_{l-1}^yS_l^zS_{l+1}^x)
	\}.
\end{eqnarray}
Applying the Jordan-Wigner transformation,
\begin{eqnarray}
S_l^x&=&1/2\Pi_{n=1}^{l-1}(1-2c_n^\dagger c_n)(c_l^\dagger+c_l);\nonumber\\
S_l^y&=&1/2i\Pi_{n=1}^{l-1}(1-2c_n^\dagger c_n)(c_l^\dagger-c_l);\nonumber\\
S_l^z&=&c_l^\dagger c_l-1/2,
\end{eqnarray}
$H$ can be rewritten as
\begin{equation}
	H=\sum_{l=1}^{N}[-J/2(c_l^\dagger c_{l+1}
	+h.c.)+J'/4i(c_l^\dagger c_{l+2}-h.c.)],
\end{equation}
which can be diagonalized by means of the
Fourier transformation,
\begin{equation}
	H=\sum_k\epsilon(k)c_k^\dagger c_k,
\end{equation}
where the energy dispersion
\begin{equation}
	\epsilon(k)=-J[\cos k-\alpha/2\sin(2k)],
\end{equation}
with $\alpha=J'/J$.

The matrix form of the Hamiltonian form can be denoted as
\begin{eqnarray}
	\rho=
	\left(
	\begin{array}{cccc}
		u_{ij} & 0 & 0 & 0\cr
		0 & \omega_{ij} & y_{ij} & 0\cr
		0 & y_{ij} & \omega_{ij} & 0\cr
		0 & 0 & 0 & u_{ij}
	\end{array}
	\right),
\end{eqnarray}
where all the elements of the matrix can
be written in terms of spin-spin correlation
functions,
\begin{eqnarray}
	u_{ij}&=&\frac14+\langle S_i^z S_j^z\rangle,\nonumber\\
	\omega_{ij}&=&\frac14-\langle S_i^z S_j^z\rangle,\nonumber\\
	y_{ij}&=&\langle S_i^x S_j^x\rangle+\langle S_i^y S_j^y\rangle,\nonumber
\end{eqnarray}
with $\langle S_i^q S_j^q\rangle~ (q=x, y, z)$ the two-point spin-spin
correlation
functions at sites $i$ and $j$, and the expectation value taken over all the
quantum states.

Using the method proposed by Lieb, Schultz,
and Mattis \cite{Lieb1961}, one can also calculate the
spin-spin correlation functions,
\begin{eqnarray}
	\langle S_l^x S_{l+m}^x\rangle=\langle S_l^y S_{l+m}^y\rangle
	=\frac{1}{4}
	\left |
	\begin{array}{cccc}
		G_{l,l+1} & G_{l,l+2} & \cdots & G_{l,l+m} \cr
		G_{l,l} & G_{l,l+1} & \cdots & G_{l,l+m-1} \cr
		\cdots & \cdots & \ddots & \cdots \cr
		G_{l,l-m+2} & G_{l,l-m+3} & \cdots & G_{l,l+1} \cr
	\end{array}
	\right|,
\end{eqnarray}
and
\begin{equation}
	\langle S_l^z S_{l+m}^z\rangle
	=-1/4(G_{l,l+m})^2,
\end{equation}
with
\begin{eqnarray}
	G_{l,l+m}=\left\{
	\begin{array}{cc}
		\frac{2}{m\pi}
		\sin(m\frac{\pi}{2}) & \alpha<1, \cr
		\frac{1}{m\pi}[1-(-1)^m]\sin(m \arcsin(1/\alpha)) & \alpha\ge1.
	\end{array}
	\right.
\end{eqnarray}

Let $t_1=4\langle S_l^x S_{l+m}^x\rangle$,
$t_2=4\langle S_l^y S_{l+m}^y\rangle$ and
$t_3=4\langle S_l^z S_{l+m}^z\rangle$.
The $3\times 3$ matrix $T$ is of the form,
\begin{eqnarray}
	T=\left(
\begin{array}{ccc}
 \frac{(t_3+1) \left(2 t_1^2+t_3-1\right)}{t_1^2-1} & 0 & 0 \\
 0 & \frac{(t_3+1) \left(2 t_1^2+t_3-1\right)}{t_1^2-1} & 0 \\
 0 & 0 & \frac{2 t_1^2+t_3-1}{t_3-1}
\end{array}
\right),
\end{eqnarray}
with eigenvalues
\begin{eqnarray}
\left\{\frac{2 t_1^2+t_3-1}{t_3-1},\frac{(t_3+1) \left(2 t_1^2+t_3-1\right)}{(t_1-1) (t_1+1)},\frac{(t_3+1) \left(2 t_1^2+t_3-1\right)}{(t_1-1) (t_1+1)}\right\}.
\end{eqnarray}

From (\ref{lqfi})  we have the LQFI,
\begin{eqnarray}
LQFI=1-\max[\frac{(t_3+1) \left(2 t_1^2+t_3-1\right)}{t_1^2-1},\frac{2 t_1^2+t_3-1}{t_3-1}].
\end{eqnarray}

By tedious calculation, we can also work out the OWQD.
The analytical expression of OWDQ for
the Heisenberg $XX$ spin chain is given by
\begin{eqnarray}
	\Delta&&=	
-2 \left(1+2t_1\right) \log \left(1+2t_1\right)-2 \left(1-2t_1\right) \log \left(1-2t_1\right)\nonumber\\
&&+\frac{1}{4}[(1-t_3+2t_1)
	\log (1-t_3+2t_1)\nonumber\\
&&+(1-t_3-2t_1)
	\log (1-t_3-2t_1)\nonumber\\
&&	+2(1+ t_3) \log (1+t_3)],
\end{eqnarray}
with the optimal value attained at $\phi=0$ and $\theta=\pi/4$.

From the above analytical expressions,
we can show the quantum fluctuations in the Heisenberg spin-$\frac12$ $XX$
spin chain system.

\begin{figure}
  \centering
  \includegraphics[width=16cm]{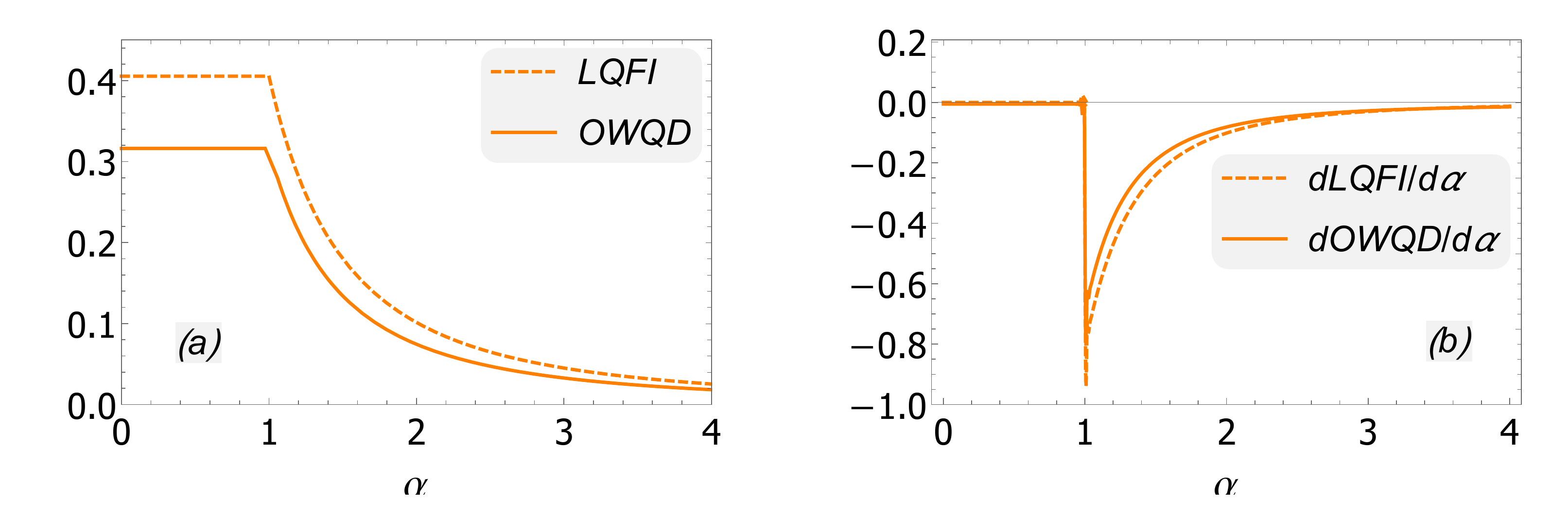}
  \caption{$m=1$: (a) LQFI and OWQD with respect to $\alpha$. The orange dashed line denotes
  LQFI. The orange solid line shows OWQD.
  (b) The derivatives of LQFI (dashed orange line) and OWQD (solid orange line) with respect to $\alpha$, respectively.}\label{f1}
\end{figure}

Fig. \ref{f1} shows the LQFI, OWQD and their
derivatives with respect to $\alpha$.
In Fig. \ref{f1}(a), the dashed line denotes
LQFI, while the solid line stands for OWQD. 
One sees that in region $[0,1]$, both
of them are in a fixed value. LQFI and
OWQD decrease quickly at first and then slowly in the region $\alpha>1$.
They have the same trends. 
However, LQFI is greater than OWQD.
They approach zero when $\alpha$ goes to infinity. 

The derivatives of LQFI and OWQD with respect to $\alpha$ are shown in Fig. \ref{f1}(b), from which
we see that the quantum phase transition happens at
$\alpha=1$. In the region $\alpha\in[0,1]$, both derivatives of LQFI and OWQD are zero.
For $\alpha>1$, the derivatives of OWQD is greater than that of
LQFI, meaning that the slope related to LQFI is steeper than to OWQD.

\begin{figure}
  \centering
  \includegraphics[width=16cm]{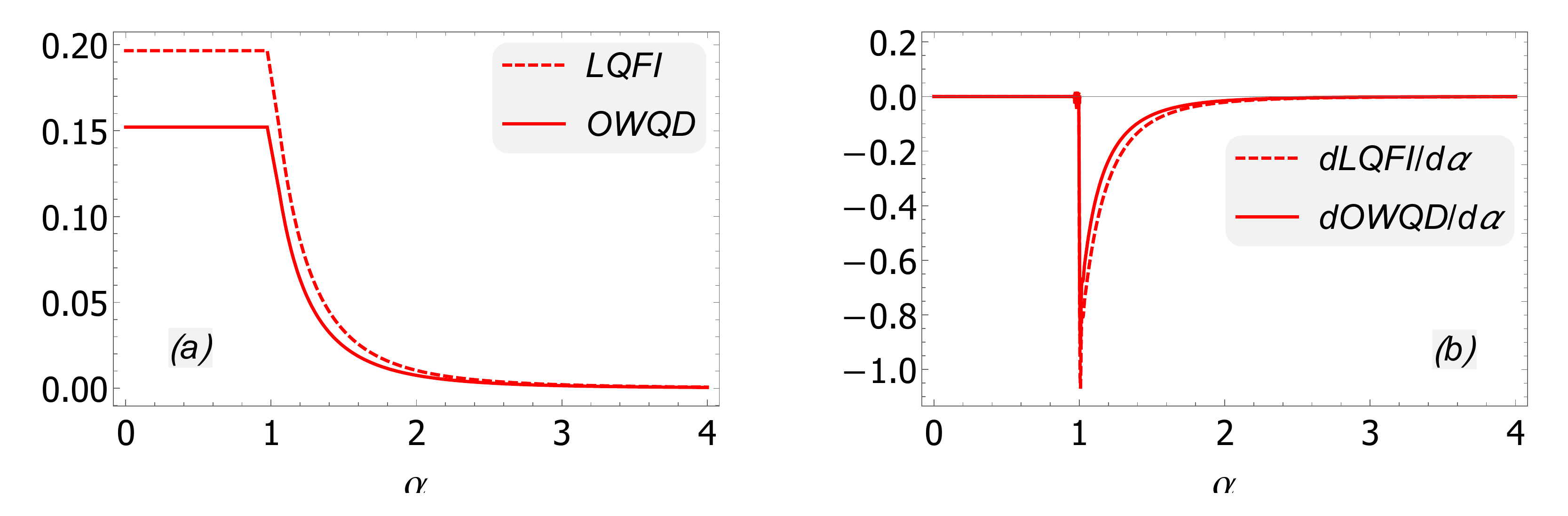}
  \caption{$m=2$: (a) LQFI and OWQD with respect to $\alpha$. 
  The orange dashed (solid) line denotes
  LQFI (OWQD). (b) Dashed (solid) orange line denotes the derivative of LQFI (OWQD) with respect to $\alpha$.}\label{f2}
\end{figure}

Fig.\ref{f2} shows LQFI and OWQD, and their derivatives with respect to
$\alpha$ when $m=2$. The LQFI and OWQD for $m=2$ are smaller
than that for $m=1$ in Fig.\ref{f2}(a), respectively.
Around the region $\alpha\in[1,2]$, the slope of the lines is more than the ones for $m=1$. 
When $\alpha$ gets larger, the LQFI is approximately coincident with the OWQD.
Both quantum correlation measures LQFI and OWQD show quantum phase transition in Fig. \ref{f2}(b).

\begin{figure}
  \centering
  \includegraphics[width=8cm]{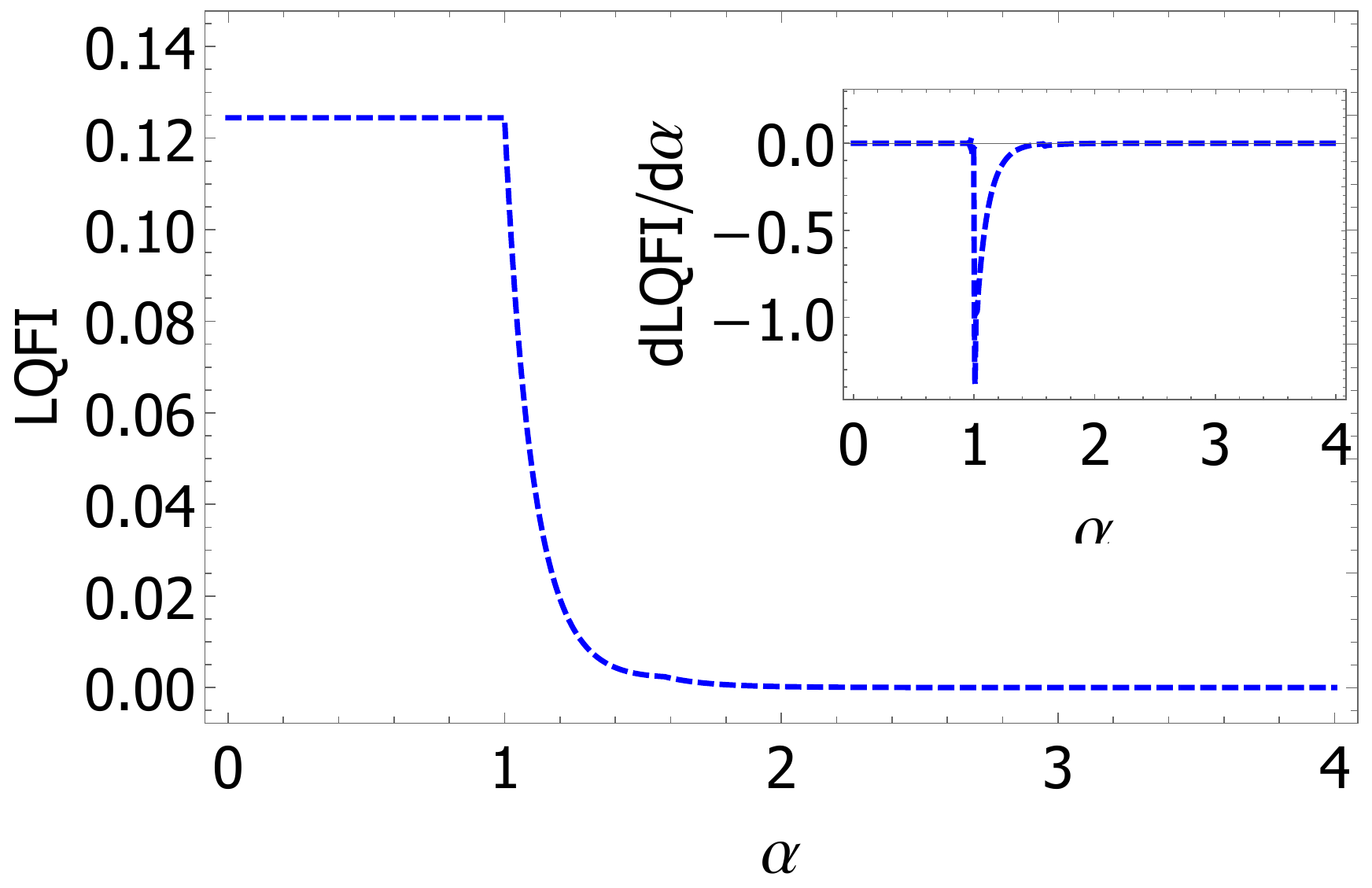}
  \caption{LQFI and its derivative (inset) with respect to $\alpha$ for $m=3$.}\label{f3}
\end{figure}

\begin{figure}
  \centering
  \includegraphics[width=8cm]{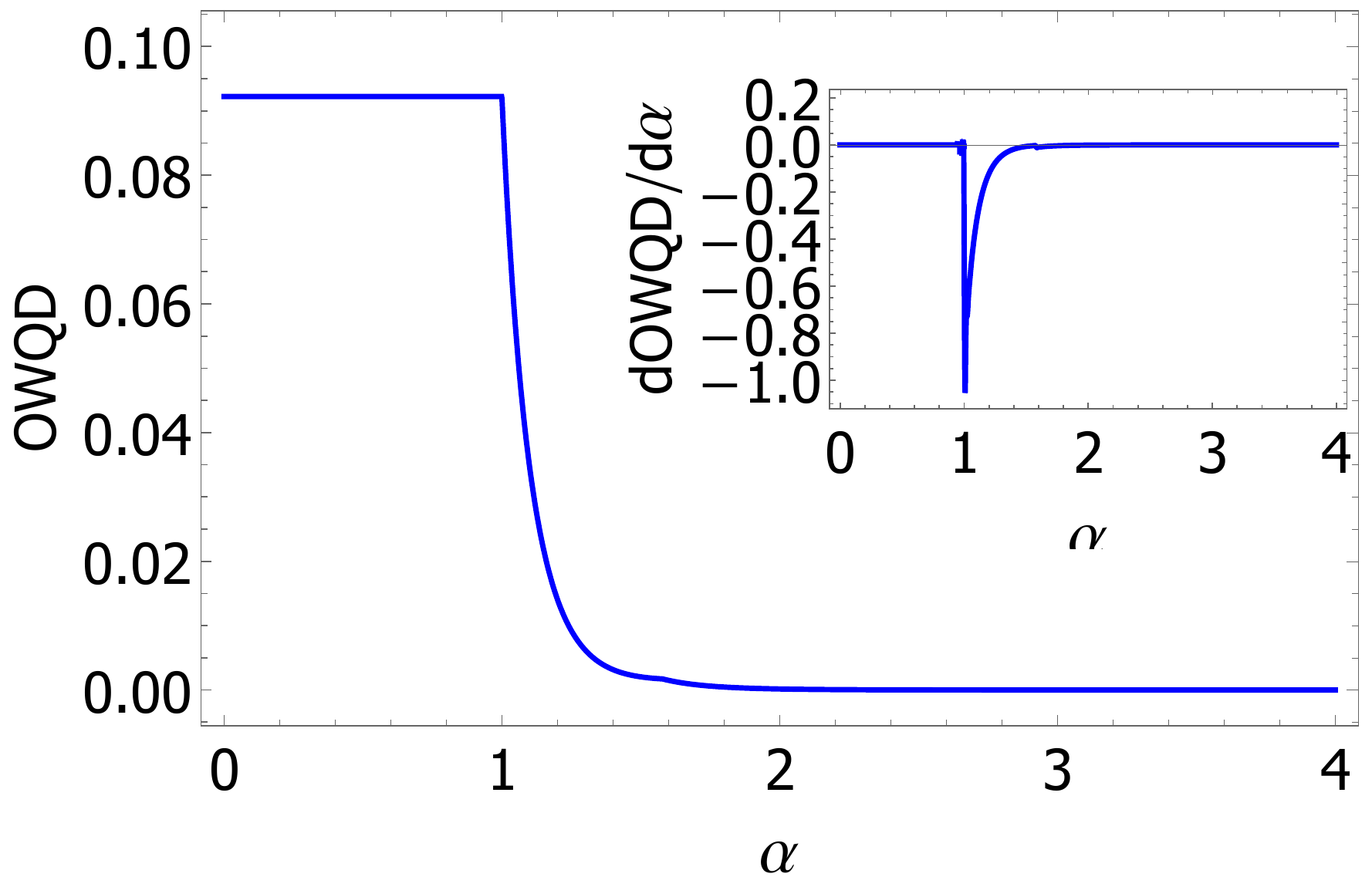}
  \caption{OWQD and its derivative (inset) with respect to $\alpha$ for $m=3$.
  The inset shows the quantum phase transition related to its derivatives.}\label{f4}
\end{figure}

Fig. \ref{f3} shows the behavior of LQFI vs $\alpha$ (dashed blue line). 
Fig. \ref{f4} shows the behavior of OWQD vs $\alpha$ (solid blue line). 
The insets show the quantum phase transition related to their derivatives.
One can see that when $m$ increases,
the slopes of the lines get larger. However,
both of them show quantum phase transition of the Heisenberg $XX$ model
by the first derivatives at $\alpha=1$.

\section{Conclusions}
We have studied the quantum phase transitions
in Heisenberg spin-$\frac12$ $XX$ spin model, showing that
the quantum phase transition happens at $\alpha=1$.
Both quantum measures, local quantum Fihser
information and one-way quantum deficit, are
able to show the quantum fluctuation and the quantum phase
transition for the Heisenberg spin-$\frac12$ $XX$ spin system. 
Our results may highlight the corresponding experimental demonstrations of
the quantum fluctuation and the quantum phase transition in the Heisenberg spin-$\frac12$ $XX$ spin systems.

\section*{Acknowledgment}
This work was supported by the Key-Area Research and Development Program of
Guangdong Province (Grant No. 2018B030326001), the NKRDP of
China (Grant No. 2016YFA0301802), the National Natural Science Foundation of China (Grant Nos. 11675113, 11765016, 11847108, and 11905131), Beijing Municipal Commission of Education (Grant No. KZ201810028042), the Beijing Natural Science Foundation (Z190005), the Natural Science Foundation of Jiangxi Province (Grant Nos. 20192BAB212005, 20192ACBL20051), and Jiangxi Education Department
Fund (Grant No. KJLD14088 and GJJ190888).


%

\end{document}